\documentclass{ere04}

\usepackage{latexsym}
\usepackage{graphicx}    % standard LaTeX graphics tool
\def\be{\begin{equation}}
\def\ee{\end{equation}}

\def\bea{\begin{eqnarray}}
\def\eea{\end{eqnarray}}
\def\ba{\begin{array}}                  %array
\def\ea{\end{array}}

\newcommand{\pa}{\partial}

\def\f{\phi}

\def\G{\Gamma}

\def\l{\lambda}

\def\m{\mu}
\def\n{\nu}

\def\p{\pi}

\def\r{\rho}

\def\t{\tau}

\def\d{\delta}
\def\dd{\textrm{d}}
\begin{document}

\title*{Propagators from characteristic surfaces}
\author{Jorge Conde
}
\institute{Instituto de F\'{\i}sica Te\'orica UAM/CSIC, C-XVI,
and  Departamento de F\'{\i}sica Te\'orica, C-XI,\\
  Universidad Aut\'onoma de Madrid 
  E-28049-Madrid, Spain.
\texttt{jorge.conde@uam.es}
}
\maketitle

We study the Goursat or characteristic problem, i.e. a hyperbolic equation with given data on a surface (the half of the standard Cauchy problem), with some kind of dimensional regularization procedure to deal with the divergences that appear. We will also comment some possible relation with a holographic setup.

\section{Introduction}
\label{sec:1}

The Cauchy problem for hyperbolic equations is well known since many years ago. We can take the wave equation in Minkowski space-time as an example. If we choose an infinite space-like surface where it is known the function and its normal derivative (with respect to the surface) the solution is fully obtained in the rest of the space. The situation becomes more delicate when the surface where we know the data is light-like, as the light cones are. The problem is then called a Goursat or characteristic one, as the surfaces are the characteristics of the equation \cite{courant} and another different procedure is needed to obtain the solution. But if we consider only a future light cone, curiously, the solution inside this  depends only on the function {\bf or} on the normal derivative of the function on the cone, but not on both. This characteristic surface is such that the derivative is not independent of the function. This is remarkable, as the causal development of a field depends in half of the data compared with the Cauchy evolution.

Although the previous considerations are purely mathematical, there might be a correspondence with physics, if we consider the holographic principle. Naively we can state that degrees of freedom in a certain space (called bulk) match one to one with the degrees of freedom on the boundary of this space\footnote{This has not to be considered as a definition of the principle, but is useful as a first hypothesis in order to make a first check of the ideas explained.}. This situation is reproduced in our simple example, where the boundary is the light cone itself, and the degrees of freedom are carried by the fields in it.

The idea of holography has been studied extensively in the last years, specially in scenarios with negative cosmological constant. The so-called $AdS/CFT$ correspondence is probably the most successful example, although the knowledge about it is still incomplete. Much less attention has been devoted to Ricci-flat geometries, with vanishing cosmological constant. In this work we will study a simple scenario of this kind and try to check a possible holographic relation. In order to do this we will need explicit classical solutions, so we will introduce a method to find them. But before we start with a description of the geometry and its properties, and how general can be this procedure.

\section{Description of the scenario}
A general geometry with some conformal properties was introduced in \cite{c2}. The line element is the following,
\be\label{ambient}
ds^2=\frac{T^2}{\ell^2}h_{ij}(\r,\vec{z})dz^i dz^j+\r dT^2+TdTd\r
\ee
with the boundary located at $\r=0$. The simplest case is to take $h_{ij}=-\d_{ij}$, then it is easy to prove that the space is $d$-dimensional flat space and that the boundary is the (origin's) future light cone. In dimension two this is called Milne universe, so we extend the definition to arbitrary dimension. Apart from this we have a set of curved geometries\footnote{By construction the geometries have vanishing Ricci tensor.}, which in terms of flat-like coordinates can be considered as deformations of Milne, always with a light-like boundary. Nevertheless we will focus on the flat case, which has the greatest degree of symmetry.

The isometries of $d$-dimensional Milne space are clearly the ones generated by $SO(1,d-1)$. The boundary is light-like, and it has a degenerate metric. Nevertheless isometries are well defined on the boundary through the Killing equations $\pounds(k) g =0$, which turn out to be the same as the (Euclidean) conformal group\footnote{There are some subtleties when $d=4$ that are irrelevant for our purposes (see \cite{goursat} for more details).}. To summarize, we deal with a bulk space with a boundary at finite (space-like) distance with conformal symmetry. Then the next step is to find how bulk fields depend on boundary data.

\section{Fields from characteristic data}
In our simple setup, the dynamics of a massless scalar field is governed by the standard wave equation. The method we present below can be generalized, first to massive fields (c.f. \cite{goursat}), and also to curved geometries with more complicated wave operators. Nevertheless for the discussion it will be enough to consider the massless scalar. Now we introduce how to obtain fields in terms of boundary data, or which is the same, the propagator from the characteristic boundary to the bulk.

\subsection{Riesz' method}
\begin{figure}
\center
\includegraphics[height=3cm]{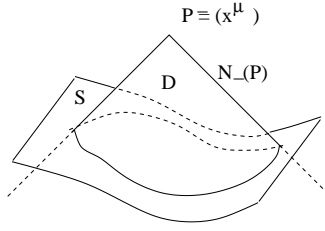}
\caption{Domain of integration for the expression (\ref{rl_d}). It is the interior of the past light cone of $P$ down to the surface $S$.}
\label{fig1}       % Give a unique label
\end{figure}

\begin{figure}
\center
\includegraphics[height=3cm]{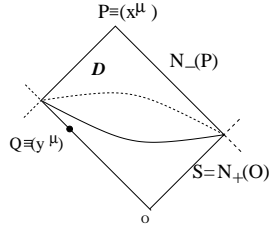}
\caption{$D$ the is the region interior to both cones, $N_-(P)$ and $N_+(O)$. We take the point $Q\equiv(y^\m)$ to be on the boundary $\pa D$. }
\label{fig2}       % Give a unique label
\end{figure}

\begin{figure}
\center
\includegraphics[height=2cm]{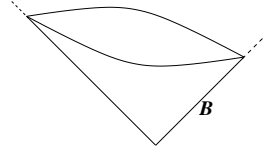}
\caption{Region of integration $B$.}
\label{fig3}       % Give a unique label
\end{figure}

This method we will use was presented by Riesz \cite{riesz} many years ago, but we are going to put it in a more modern way and try to relate it to well known techniques such as dimensional regularization. It is based on the Riemann-Lioville integral
\be
I^{\l} f(x)=\frac{1}{\G (\l)}\int^{x}_{a}{dt f(t)(x-t)^{\l-1} }
\ee
In principle this is well defined for $\l > 0$ ($f$ vanishes for $x=a$), but noticing that
\be\label{cont_anal}
\begin{array}{l}
I^{\n}I^{\l}f=I^{\n+\l}f \\
I^{0}f=f
\end{array}
\ee
 the expression can be continued to negatives values of $\l$ (basically the Gamma function cancels the integral divergences). This construction can be easily generalized to higher dimensions by the formula
\be\label{rl_d}
I^{\l}\f(x)=\frac{1}{H_d(\l)}\int_{D}{d^{d-1} y\ \f(y)\ \t_{xy}^{\l-d}}
\ee
where $\t_{xy}$ is the geodesic distance between the points $P$ and any point in the domain $D$ (see figure \ref{fig1}).
If we take a definite value for the pre-factor $H$ such as
\be
H_d(\l)=\p^{\frac{d}{2}-1}2^{\l-1}\G\left(\frac{\l}{2}\right)\G\left(\frac{\l+2-d}{2}\right)
\ee
the expression satisfies the desired properties (\ref{cont_anal}) for the right analytic continuation. Moreover, it is also true that
\be\label{box}
\Box I^{\l+2}=I^{\l}
\ee
so we can identify $\Box=I^{-2}$, where $\Box$ is the D'Alambertian. With this in mind we take the surface $S$ to be the origin's future light cone, which is not a space-like surface but a light-like one. The domain is presented in figure \ref{fig2}.
Now integration on $D$ is related to integration over $\pa D$ thanks to Stokes' Theorem, for instance it is true that \footnote{We use ``$\dd$'' for the exterior derivative and ``$*$'' for the Hodge dual.}
\be
\int_{D}{(f\ \dd *\dd g\ -\ \dd *\dd f\ g)}=\int_{\pa D}{(f\ *\dd g\ -\ *\dd f\ g)}
\ee
Let us now consider the functions $f=\f(x)$, the solution of the wave equation \newline $\Box\f(x)=\dd*\dd\f(x)=0$ with prescribed value $\f|_{N_+}=\psi$, and $g={\tau_{xy}}^{2\l}$. This latter choice leads, using properties (\ref{cont_anal}) and (\ref{box}), to the desired solution
\be
\f(x)=\lim_{\l\to1-\frac{d}{2}}\frac{1}{H_d(\l)}\int_{\pa D}{(\f\ *\dd\tau^{2\l}-\tau^{2\l} *\dd\f)}
\ee
This expression has several relevant features. The factor $H_d$ has a divergent contribution, $\G\left(\frac{-d}{2} \right)$, which divergences from integration will cancel. Moreover, as the analytic continuation parameter $\l$, when we take the appropriate limit, depends on the space-time dimension $d$, we can regard the expression as formally dimensionally continued, and then use the well known results of dimensional regularization in order to integrate. The last point to note is that the presence of $\tau$ in the integrand makes it vanishing on $N_-(P)$, as this is the past light cone of $P$. Then the domain of integration is restricted to the origin's light cone (the surface with given values for the field $\f$) up to the intersection with $N_-(P)$, which we call $B$ (see figure (\ref{fig3})).

\subsection{Explicit solution}
From the previous subsection we know the wave equation solution for a scalar field with prescribed values (for the field only) on the origin's future light cone, which is a characteristic surface.
\be\label{general}
\f(x)=\frac{1}{H_d(1-\frac{d}{2})}\int_{B}{(\phi(y)\ *\dd{\tau_{xy}}^{2-d} - {\tau_{xy}}^{2-d}\ *\dd\f(y)) }
\ee
where we have to understand $d$ as a complex parameter. Now we have to choose explicit coordinates. In order to let the boundary dependence in term of dummy variables we shall take standard flat coordinates, $x\equiv(t,\vec{x})$, and $y\equiv(t',\vec{y}')$. The surface $B$ is parametrized by $d-1$ variables, for instance  $t'$ and $\vec{n}'=\frac{\vec{y}'}{|\vec{y}'|}$. We also have to remember that the field's normal derivative on the boundary depends on the field itself. This translates to the formula in the way that the second contribution of (\ref{general}) can be integrated by parts (and as $\pa D$ has no boundary there is only one contribution after integration). With all this the solution is the following.
\be\label{solucion}
\f(t,\vec{x})=\frac{1}{H_d(1-\frac{d}{2})}\int{d\vec{n}'}\int^{\infty}_0{dt'\ t'^{d-3}\ \left( s^2-2t'(\vec{n}'\vec{x}-t) \right)^{-\frac{d}{2}}\ H\left(\frac{s^2}{2(\vec{n}'\vec{x}-t)}\right)\ \psi(t',\vec{n}')}
\ee
where $s^2=t^2-\vec{x}^2$, $H$ is a Heaviside step function (as the integration domain is finite) and $\psi$ is the arbitrary boundary value. This expression is written as
\be
\f(x)=\int_{N_+}{K(x,y)\ \psi(y)\ dy}
\ee
although the kernel $K$ has to be taken with care as it depends on a step function and a complex parameter, which can only be regarded as integer {\bf after} integration. This expression is a solution of $\Box\f=0$, and the kernel tends to a delta function as the point $P$ tends to the boundary \footnote{Perhaps the simplest way in order to check this is to consider the boundary value $\psi$ as an analytic function and expand it as a power series, $\psi(t',\vec{n}')=\sum_{m}{t'^{a_m} \varphi_m(\vec{n}')}$, then integrate on $t$ and let $t\to \vec{x}$.}, so it reduces to the prescribed value on the boundary.

\subsection{A theory on the boundary?}
Once we have under control fields in Milne space we ask ourselves about a possible holographic interpretation in the same fashion as $AdS/CFT$ (see \cite{witten}), that is, if the action (in this case only the contribution of a scalar field) of a physical theory in Milne geometry can be re-interpreted in terms of a theory living on the boundary. 

The simplest approximation, although with all the desired properties, is to consider the on-shell action for the bulk theory, 
\be\label{accion}
\mathcal{S}|_{on-shell}=\int_{M}{ \dd \f\ \wedge * \dd \f }=\int_{\pa M}{\f\ * \dd \f }
\ee
where the final expression comes from integration by parts, and it depends only on boundary values. The field $\f$ is the one given by the solution (\ref{solucion}), then we only have to obtain the field normal derivative $*\dd \f$ and let it tend to the boundary value. If we remember that the kernel $K$ of the previous subsection tends to a delta function near the boundary, its normal derivative is related basically to the first derivative of the delta, so the final result we get in terms of the boundary value $\psi$ after some manipulation is of the kind
\be\label{rfinal}
\mathcal{S}|_{on-shell}\sim \int{dx\ \psi(x)}\int{dx'\ \frac{1}{x}\ \d(x-x')\ \psi(x')}\sim\int{dx\ \frac{1}{x}\ \psi(x)^2}
\ee
Clearly this happens to be a boundary action without physical meaning, so it seems not to exist a holographic interpretation. The main difference with the $AdS/CFT$ result is that the $AdS$ boundary is at infinite distance, and in fact it is not a true boundary but a conformal one. This is crucial because in the action appears a contribution due to the integration measure, and this combined with the normal derivative gives a result different from the delta in (\ref{rfinal}), and then the action can be interpreted as a generator of Green functions.

\section{Conclusions}

We have presented an old method to solve Goursat or characteristic (and Cauchy, as the construction is quite general) problems from a more modern point of view. The method allows us to calculate propagators in some simple scenarios and try to verify the possibility of a holographic relationship between theories living in a space with vanishing cosmological constant and in its boundary. The result is similar to the one in $AdS/CFT$ except for a (crucial) contribution that comes from the location of the boundary. In the $AdS$ case the metric blows up on the boundary, as this is at Penrose infinity, but in our setup the metric is smooth everywhere, and then it does not kill the delta behaviour in (\ref{rfinal}).

\section*{Acknowledgements}
I am grateful to Enrique \'Alvarez and Juan F. G. Cascales for useful comments on the manuscript.

%\printindex
\end{document}